# Requirements, Formal Verification and Model transformations of an Agent-based System: A CASE STUDY

Nadeem Akhtar (Corresponding author)

PhD - IRISA – University of South Brittany - FRANCE

Department of Computer Science and IT, The Islamia University of Bahawalpur

Baghdad-ul-Jadeed campus, Pakistan

Tel: +92-331-2116491, E-mail: nadeem.akhtar@iub.edu.pk

*The research is financed by Department of Computer Science and IT, The Islamia University of Bahawalpur, Pakistan*

**Abstract**

One of the most challenging tasks in software specifications engineering for a multi-agent system is to ensure correctness. As these systems have high concurrency, often have dynamic environments, the formal specification and verification of these systems along with step-wise refinement from abstract to concrete concepts play major role in system correctness. Our objectives are the formal specification, analysis with respect to functional as well as non-functional properties by step-wise refinement from abstract to concrete specifications and then formal verification of these specifications. A multi-agent system is concurrent system with processes working in parallel with synchronization between them. We have worked on Gaia multi-agent method along with finite state process based finite automata techniques and as a result we have defined the formal specifications of our system, checked the correctness and verified all possible flow of concurrent executions of these specifications. Our contribution consists in transforming requirement specifications based on organizational abstractions into executable formal verification specifications based on finite automata. We have considered a case study of our multi-agent system to exemplify formal specifications and verification.

## 1. Introduction

One of the most challenging tasks in software engineering for multi-agent system is to ensure continuous correctness, especially as software systems are increasingly used in dynamic and often distributed environments, to support the formal specification of software systems whose architecture can change, to support their analysis with respect to functional as well non-functional properties, to support their design by step-wise refinement from abstract to concrete specifications and full code generation, as well as their subsequent evolution.

An agent is a computer system situated in some environment, capable of autonomous actions in this environment in order to meet its design objectives [Wooldridge and Jennings, 1995]. Multiple agents are necessary to solve a problem, especially when the problem involves distributed data, knowledge, or control. A multi-agent system is a collection of several interacting agents in which each agent has incomplete information or capabilities for solving the problem [Jennings et al., 1998]. These are complex systems and their specifications involve many levels of abstractions. While considering multi-agent methods we have selected Gaia [Zambonelli et al., 2003] [Wooldridge et al., 2000] as it is based on computation organization with various interacting roles. It is both general, in that it is applicable to a wide range of multi-agent systems, and comprehensive, in that it deals with both the macro-level (societal) and the micro-level (agent) aspects of systems. It has a concrete syntax to express properties, and it is suitable to model behaviors.

Gaia multi-agent method is the principal candidate for the specifications whereas Finite State Process (FSP) with Labelled Transition System Analyzer (LTSA) [Magee and Kramer, 2006] has been selected for the verification of specifications as it is a formal language specifically useful for specifying concurrent behavior and can generate finite automates by using LTSA.

We have considered a case study consisting of small multi-agent robotics software agents working in a closed environment, formal methods are used i.e. Gaia specifications and then it's FSP modelling for the formal specification and verification of the system.

In this article the main focus is on the formal specification and property-checking aspects of the system. Section-2 presents the problem statement, section-3 the Background studies, section-4 the requirements and formal verification of our case study, section-5 presents the transformation from requirement specifications to formal verification, section-6 put forward the lessons learned and conclusion.

## 2. Problem statement

Multi-agent systems are specialized systems with greater autonomy, complexity, abstraction and concurrency and therefore we need specialized methods and formal languages.

The objectives to have a formal foundation for the languages and tools are: to improve understanding of specifications, to enable rigorous analysis of the system properties, to be as certain as possible that the transformations and implementation are property-preserving and error-free, to improve the quality of the whole development process, and to provide a firm foundation during the adaptation and evolution process.

There is a need of formal methods, techniques, design tools and languages for specification definition, architecture description and definition of layers of abstractions. There are some existing multi-agent methods such as Gaia and TROPOS [Giunchiglia et al., 2002] and among these method we have found Gaia as the most suitable one for our work as it recognizes the organizational structure is a primary dimension for the development of agent system and an appropriate choice of it is needed to meet both functional and non-functional requirements. Presently we have proposed an architecture consisting of formal methods and languages for requirement definition, specification definition and formal verification.

It has been observed that there is a need of formal verification for Gaia specifications of our system. Here we start from Gaia based requirement specifications, Finite state machines based formal verification and then a MRDS (Microsoft Robotics Developer Studio) [MRDC, 2007] based simulation implementation.

## 3. Background studies

### *3.1 Formal methods*

Formal methods are based on solid mathematical foundations. Formal verification is the act of proving or disproving the correctness of underlying system algorithms with respect to certain formal specifications using formal methods of mathematics.

Two main types of formal methods are available: algebraic approaches and model-checking. Algebraic approaches such as B [Abrial, 1996] describe a system with axioms and then prove a property on the specification as a theorem to be demonstrated from these axioms. However theorem provers that are required to elaborate the proof are difficult to use and still require highly skilled and experience engineers.

In contrast, model-checking [Clarke et al., 2000] [Berard et al., 2001] is the exhaustive investigation of a system state space. This technique is limited by the combinatorial explosion and can mainly address finite systems. However, recent symbolic techniques scale up to more complex systems. Here by complex we mean a system with a large number of independent interacting components, with nonlinear aggregate activity, with concurrency between components and constant evolution. Thus, since formal verification techniques are getting more mature, our capability to build even more complex systems also grows quickly.

Formal verification can achieve complete exhaustive coverage of the system thus ensuring that undetected failures in the behavior are excluded. The goal is to formally verify that each component is consistent with the rest; agents are able to fulfill their commitments. Correctness of an agent software system can be proved by formalizing different components and processes in the life-cycle. The case study is presented in section-6 with the formal correctness which requires formal verification.

### *3.2 Gaia Overview*

Gaia [Zambonelli et al., 2003] [Wooldridge et al., 2000] clearly identifies the appropriate organizational abstractions in a multi-agent system and details the analysis and design of such systems. These organizational abstractions are necessary to design and built complex systems.

Gaia is not directly related with particular modelling techniques and implementation issues. After the successful completion of the design process, developers are provided with a well-defined set of agent roles to manually implement and instantiate, according to the defined agent and services model.

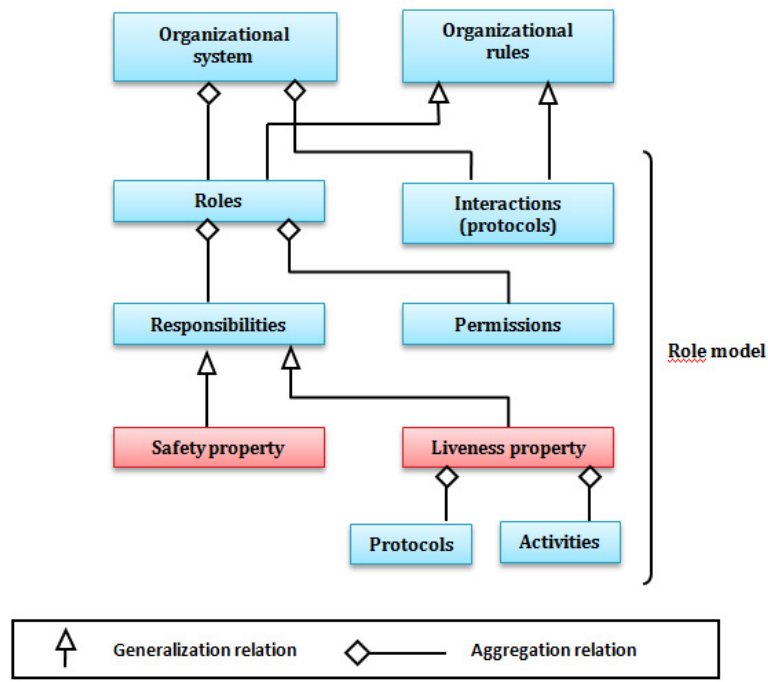

Figure 1. Gaia's organizational structure (Responsibilities, Permissions, Activities, and Protocols)

As shown in figure-1 Gaia method has an organizational structure based on roles and interactions, Agent roles have Responsibilities, Permissions, and further responsibilities are classified into safety and liveness properties. These safety and liveness properties are important and can be extended and formally verified by using formal methods and techniques.

Agent, services and acquaintance model constitute the design phase.

The Gaia's role model manages permissions and responsibilities.

Permissions are classified into two types internal and external. Internal permissions are those resources or properties associated with only the agent and don't involve interaction with the environment or another agent. Permissions are mainly aimed at

1) Identifying the resources that can legitimately be used to carry out the role i.e. what can be spent while carrying out the role
2) In order to carry out a role, an agent will typically have to access environmental resources and possibly change.

Whereas responsibilities attributes determine the expected behavior of a role and the key attributes associated with a role. They include the liveness and safety properties that play important role for the formal verification of the system.

The key phases of Gaia can be summarized as:

- The definition of the system's organizational structure in terms of its topology and control regime. It involves considering: (i) organizational efficiency (ii) real-world organization and (iii) need to enforce the organizational rules.
- Identify the roles in the system. They will typically correspond to individual agents, either within an organization of agents or acting independently

    Output: a prototypical roles model, a list of the key roles that occur in the system, each with an informal description that is not elaborated.

- For each role, identify and document the associated protocols

    Output: an interaction model, which captures the recurring patterns of role interactions.

- Using the interaction model (protocol definitions) as a base to elaborate the roles model

    Output: a fully elaborated roles model, which documents the key roles occurring in the system, their permissions and responsibilities, together with the protocols and activities in which they participate.

The completion of the role and interaction models based on the adopted organizational structure and involves separating whenever possible the organizational-independent aspects from the organizational-dependent ones. This demarcation promotes a design-for-change perspective by separating the structure of the system from its goals.

**Safety and liveness properties**

Safety property is an invariant which asserts that "nothing bad happens"', that is an acceptable state of affairs is maintained. Safety property P = {a1, a2, ..., an} defines a deterministic process that asserts that any trace including actions in the alphabet of P, is accepted by P.

Liveness property asserts that ``something good happens'' that describe the states of system that an agent must bring about given certain conditions. In the Gaia's role model liveness properties are specified via liveness expressions which defines the potential execution trajectories through the atomic components (activities and interactions) associated with the role.

An activity corresponds to a unit of action that the agent may perform, which does not involve interaction with any other agent. Protocols, on the other hand are activities that require interaction with other agents [Zambonelli et al., 2003].

Table 1. The capitals, assets and revenue in listed banks

| Operator | Interpretation |
|---|---|
| x.y | x followed by y |
| x\|y | x or y occurs |
| x* | x occurs 0 or more times |
| $x^+$ | x occurs 1 or more times |
| $x^\omega$ | x occurs indefinitely often |
| [x] | x is optional |
| x\|\|y | x and y interleaved |

The most widely used formalism for specifying liveness and safety properties is temporal logic, and the use of such formalism has been strongly advocated for use in agent systems [Wooldridge, 2000].

*3.3 FSP and LTS Overview*

FSP is a process algebra notation in the form of finite state processes used for the concise description of component behavior particularly for the concurrent systems. It provides the means to formalize specification of software components and architecture. Each component consists of processes and each process has finite number of states and is composed of one or more actions. Because of its strong parallel constructs it is used in particular for parallel and concurrent systems. There exists concurrency between elementary calculator activities for which there is a need to manage the interactions, communication and synchronization between processes.

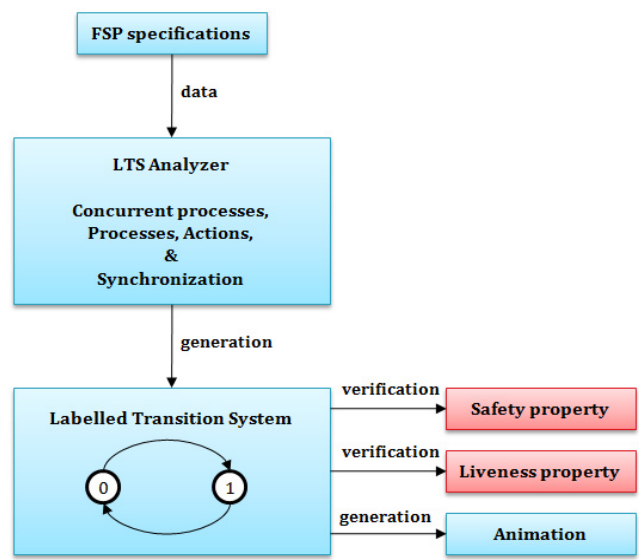

Figure 2. FSP and the Analysis toolkit LTSA

[Magee and Kramer, 2006] have proposed an analyzer LTSA. A system in the LTSA is modelled as a set of interacting finite state machines along with their properties. LTSA performs compositional reachability analysis to exhaustively search for violations of the desired properties. More formally, each component of a specification is described as LTS, which have all the possible states a component may reach and all the possible transitions it may perform. However, explicit description of an LTS in terms of its states; set of action labels and transition relation is cumbersome for other than small systems.

As shown in figure-2 FSP specify the behavior of concurrent systems to the LTSA which in turn generates finite automata based on Labelled Transition Systems and makes it possible to view the graphical representation of the LTS corresponding to FSP specifications. Therefore FSP is used to formalize specification of software components and Labelled Transition System is used to verify system-level concurrency properties. As a result we are able to obtain a concurrent system in which there are processes working in parallel and there are synchronizations between different processes.

**Progress and Deadlock detection**

The regular occurrence of some actions in a system execution indicates that system behavior progresses as desired or expected. In the context of an infinite execution, regularly means infinitely often i.e. a property that asserts that an action is expected to occur infinitely often in every infinite execution of the system, the liveness properties of this type are progress.

Progress P = {a1, a2 … an} {defines a progress property P which asserts that in an infinite execution of a target system, at least one of the actions a1, a2, ... , an will be executed infinitely often [Magee and Kramer, 2006].

In systems with parallel processes deadlock refers to a situation where two or more processes are unable to proceed because each is waiting for one of the others to do something. In these systems one of the major issues is the deadlock detection and prevention to conceptualize a system that is deadlock free. In LTS a system having two or more parallel processes, each having shared actions i.e. having actions between different processes being synchronized by renaming there exists significant possibilities of deadlock. Progress and deadlock are two terms that are somewhat related if there is a deadlock in the system than it results into a progress violation.

## 4. CASE STUDY: multi-agent robotics transport system

In this section we present our multi-agent robotic system which is composed of transporting agents. Overall mission is to transport material from one storehouse to another. They move in their environment which is static i.e. the topology of the system does not evolve at run time. There is a possibility of collision between agents while transportation. Collision resolution techniques are applied to avoid system deadlock.

### *4.1 Types of Agents*

There are three types of agents
1. Carrier agent: agent that transports material from one storehouse to another one, it can be loaded or unloaded and can move both forward and backward direction. Each road section is marked by a sign number and the carrier agent can read this number. It also has a sensor to detect a collision with another carrier.
2. Loader / Un-loader agent: It receives/delivers material from the storehouse, can detect if a carrier is waiting (for loading or unloading) by reading the presence sensor, it ensures that the carrier waiting to be loaded is loaded and the carrier waiting to be unloaded is unloaded.
3. Store-manager agent: manages the material count in the storehouse and transports the material between the storehouse and the loader/un-loader.

### *4.2 Environment*

There is a road between storehouse-A and storehouse-B which is composed of a sequence of interconnected sections of fixed length as shown by figure-3. Each road section has a numbered sign, which is readable by carrier agents.

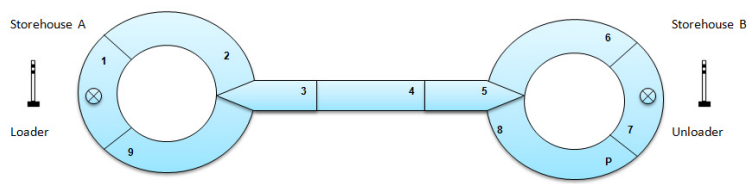

There are three types of road sections depending upon the topology of the road. Each of the three types of road sections has a unique numbered sign. The road is single lane and there is a possibility of collision between agents. There is a roundabout at storehouse-A and storehouse-B.

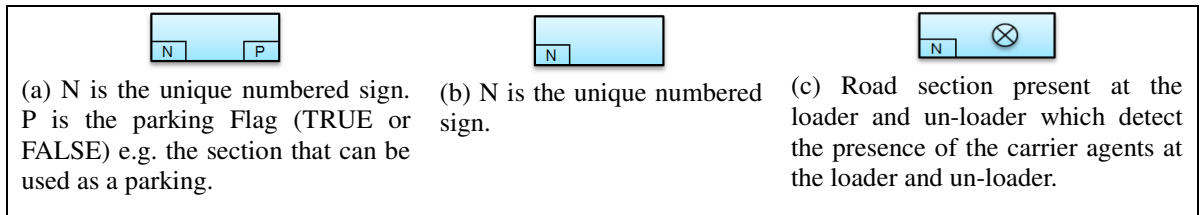

(a) N is the unique numbered sign. P is the parking Flag (TRUE or FALSE) e.g. the section that can be used as a parking.

(b) N is the unique numbered sign.

(c) Road section present at the loader and un-loader which detect the presence of the carrier agents at the loader and un-loader.

Figure 3. Environment with road partitions

*4.3 Scenario*

In the case study we have used a particular road topology consisting of nine road partitions as shown in figure-3. It's a mini-route that can be presented here in this paper with all its states.

We have considered the case in which initially storehouse A is full and storehouse B is empty. The main task of the carrier is to transport the material from storehouse A to storehouse B until the storehouse A is empty. Loader at the storehouse A loads the carrier with material and the Un-loader at the storehouse B unloads the carrier. The store-manager keeps a count of material in each storehouse. In this case the environment is static.

At the central section (3, 4, 5) there is a possibility of collision between carrier agents coming from the opposite directions. Priority is given to the loaded carriers i.e. if there is a collision between a loaded and an empty carrier than the empty carrier moves back and waits at the parking region of the route until the loaded carrier passes and unloads. The parking region as shown in the figure consists of the road partition 8.

*4.4 Gaia development and transformation*

The major part of the work is to take the Gaia specifications and then use them in a way that they can be verified by using FSP language. Gaia method as described in section-4 consists of a number of models; we may be looking into only the roles model and interaction model which constitutes the analysis phase of Gaia.

**4.4.1 Role model**

The role of an agent defines what it is expected to do in the organization, both in concert with other agents and in respect of the organization itself. Often an agent's role is simply defined in terms of the specific task that it has to accomplish in the context of the overall organization. Organizational role models precisely describe all the roles that constitute the computational organization. They do this in terms of their functionalities, activities, responsibilities as well as in terms of their interaction protocols and patterns. In the role model the liveness and safety expressions play important role for system verification.

In our system for the carrier agent there are Move_full and Move_empty roles. These roles are better adapted for this type of route where priority is given to the loaded carriers. Here in this paper due to space constraints we present the Move_full role of our system i.e. role of a loaded carrier agent moving from Storehouse-A to Storehouse-B.

Table 2. Move_full role

| |
|---|
| Role Schema: **Move_full** |
| **Description:**<br>Role of a loaded carrier moving from storehouse A to storehouse B. |
| **Protocols and Activities:**<br>readSign, movetoNext, collisionSensorTrue, carrierWait, readUnloadSign, waitforUnloading, unloadCarrier |
| **Permissions:**<br>reads:     *sign_number (external)*<br>           *collision_sensor (internal)*<br>changes:  *position (internal)*<br>           *next_position (external)* /// (True or False) checks if the next position is available |

| Role Schema: **Move_full** |
|---|
| **Responsibilities:** <br> *Liveness:* <br> **Move_full** = *Move*.(<u>readUnloadSign</u>.waitForUnloading.unloadCarrier) <br> *Move* = (<u>readSign</u>. <u>movetoNext</u>)+ <br>      | (<u>collisionSensorTrue</u>.*Wait*).(<u>readSign</u>.<u>movetoNext</u>)+ <br> *Wait* = <u>carrierWait</u>+ <br> *Safety*: <br> is_Full(c) ∧ can_movetoNext(sn) <br> where c is for carrier and sn for the sign number |

Here activities (underlined) are ReadSign, MovetoNext, CollisionSensorTrue, CarrierWait, and ReadUnloadSign. And there are two protocols WaitforUnloading and UnloadCarrier WaitforUnloading: when a loaded carrier reads the unload sign i.e. it reaches the unload road partition, it stops there and waits until it is unloaded.

Consider the Liveness property of the Move_full role. It shows all the activities and protocols that make up the role. The carrier has two choices, first it can read sign and move to the next road partition, second it detects the collision sensor then it waits, at the end it reads the unload sign i.e. at the road partition in front of the un-loader, and in this case the carrier stops and waits for being unloaded, so now it's no more a loaded carrier and is no more part of the Move_full role. The safety property is an invariant which states that any carrier playing that role schema is currently loaded. Here next_position identifies the direction of the loaded carrier at the route.

**4.4.2 Interaction model**

There are dependencies and relationships between the various roles in a multi-agent organization which are the set of protocol definitions, one for each type of inter-role interaction. Here table-3 shows the protocol definitions related to Move_full and Move_empty role.

Table 3. Protocol definitions related to Move_full and Move_empty role

**Move_full role protocols**

| waitForUnloading ||
|---|---|
| Move_full | Unload |
| The full carrier agent waits for the un-loader agent ||

*sign_number*

*position*

| unloadCarrier ||
|---|---|
| Move_full | Unload |
| The full carrier agent is unloaded by the un-loader agent ||

*sign_number*

*position*

**Move_empty role protocols**

| waitForLoading ||
|---|---|
| Move_empty | Load |
| The empty carrier agent waits for the loader agent ||

*sign_number*

*position*

| loadCarrier ||
|---|---|
| Move_empty | Load |
| The empty carrier agent is loaded by the loader agent ||

*sign_number*

*position*

### *4.5 FSP specifications of the system*

We formally defined our system using FSP, verified all the possible flow of executions. In our system the road is the environment and each carrier has its particular route. The route is the path taken by the carrier agents on the road to transfer the material from one storehouse to another. Here below are the FSP specifications for the route.

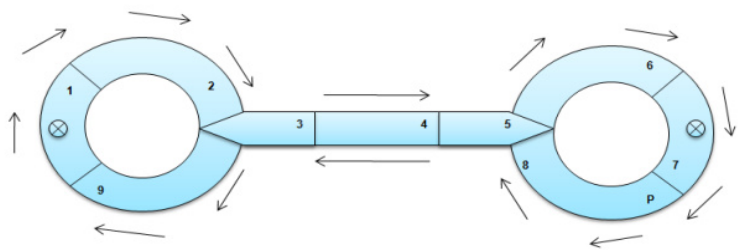

```
1   range R = 1..9
2   ROUTE = EMPTY_ROUTE[9],
3   FULL_ROUTE[v:R]=(
4           when (v==7)   readunloadSign   -> FULL_ROUTE[v]
5         | when (v!=7)   readSign[v]      -> FULL_ROUTE[v]
6         | when (v>=1&v<=6)movetonext     -> FULL_ROUTE[v+1]
7         | when (v==7)   waitforunloading -> EMPTY_ROUTE[7]
8   ),
9   EMPTY_ROUTE[v:R]=(
10          when (v==1)   readloadSign     -> EMPTY_ROUTE[1]
11        | when (v!=1)   readSign[v]      -> EMPTY_ROUTE[v]
12        | when (v==7)   movetonext       -> EMPTY_ROUTE[v+1]
13        | when (v==8)   movetonext       -> EMPTY_ROUTE[5]
14        | when (v==5)   movetonext       -> EMPTY_ROUTE[v-1]
15        | when (v==4)   movetonext       -> EMPTY_ROUTE[v-1]
16        | when (v==3)   movetonext       -> EMPTY_ROUTE[9]
17        | when (v==9)   movetonext       -> EMPTY_ROUTE[1]
18        | when (v==3)   movetoprevious   -> EMPTY_ROUTE[v+1]
19        | when (v==4)   movetoprevious   -> EMPTY_ROUTE[v+1]
20        | when (v==5)   movetoprevious   -> EMPTY_ROUTE[8]
21        | when (v==1)   waitforloading   -> FULL_ROUTE[1]
22  ).
```

With the help of the LTSA this code generates a finite automate of the route with all the possible executions that can take place on this route. Carrier moves in a clockwise direction on this route. Here the route has been classified in two types the FULL_ROUTE (i.e. path taken by full carrier) and the EMPTY_ROUTE (i.e. path taken by empty carrier).

The LTS automate generated by the given FSP specifications has 14 states, so we have broken them into two parts to represented here as shown in figure-4. As priority is for the full carriers so in case of collision between full and empty carrier parking section-8 is used by the empty carrier to park there and leave the central portion of the route open to the full carrier. As shown by the FSP specifications, the carrier performs certain actions as readSign, movetoNext and movetoPrevious.

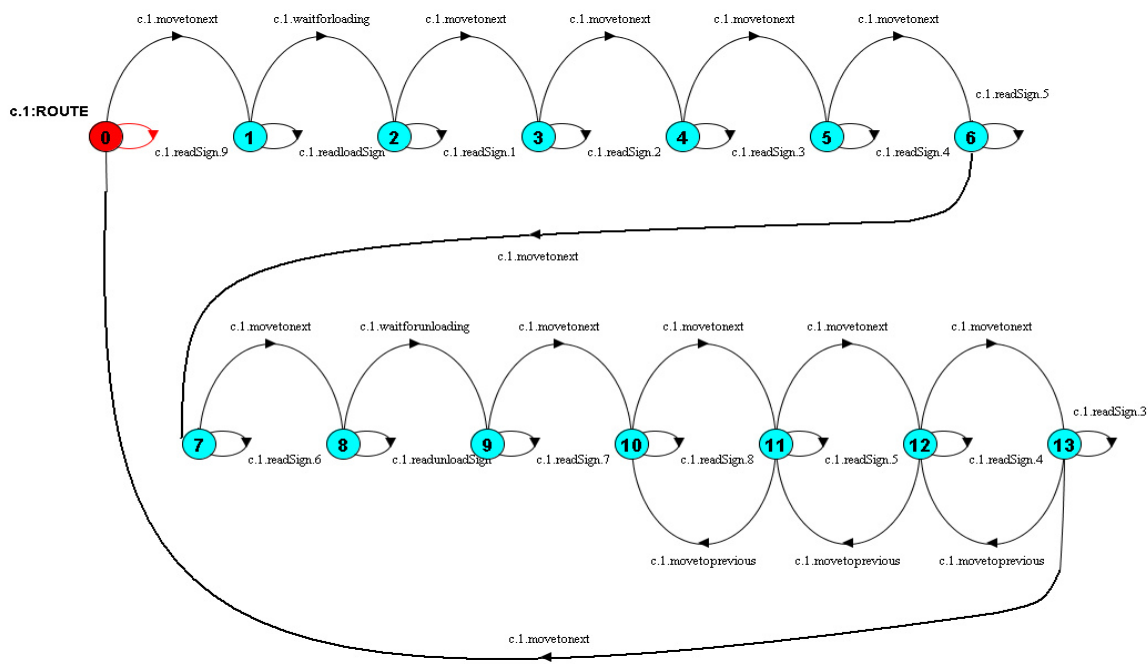

Figure 4. LTS specifications for mini-route along with LTS

Section 1 is the loader road section i.e. the road section in front of the loader and section 7 is the un-loader road section i.e. the road section in front of the un-loader. The central portion road sections 3,4,5 are shared by both full and empty carriers, so there exists a possibility of collision between full and empty carriers, to avoid system deadlock after collision we have taken some collision resolution steps. The move_full role of the carrier agent is formally specified in the form of FSP for formal verification as shown below.

```
const Max = 9 // road partitions
range R = 1..Max // range of road partitions {1..9}
MOVE_FULL(S=1) = MOVEFULL[S],
MOVEFULL[s:R] = ( readSign[s] -> MOVEFULL[s]
   | when(s>=1 && s<7) movetonext[s] -> MOVEFULL[s+1]
   | when(s>=3 && s<=5) collisionSensorTrue[s] -> COLLISION
   | when(s==7)
      readUnloadSign -> waitforUnloading -> unloadCarrier -> gotoMoveempty ->
Stop),
COLLISION = (collisionSensorTrue -> Wait),
Wait = (carrierWait -> Stop).
```

The LTS generated by the above FSP with all its states are not represented because of space constraints.

The transfer of material from one storehouse A to another storehouse B can be specified by the FSP specifications along with the LTS as shown in figure-5. In this case the maximum stock is set to 2 specify the system with a minimize state LTS that can be presented here in this paper.

```
const MaxS = 2        /// maximum number of Stock
range S = 0..MaxS
STOCKFULL_MANAGEMENT = STOCK_FULL[MaxS],
STOCK_FULL[st:S] = ( stockCountA[st] -> STOCK_FULL[st]
   | when(st>0)   decrementStockA -> send -> STOCK_FULL[st-1]
   | when(st==0)  stockEmptyA -> STOP).

STOCKEMPTY_MANAGEMENT = STOCK_EMPTY[0],
STOCK_EMPTY[st:S] = ( stockCountB[st] -> STOCK_EMPTY[st]
   | when(st<MaxS)   receive -> incrementStockB -> STOCK_EMPTY[st+1]
   | when(st>=MaxS)  stockFullB -> STOP).

||STOCKSYSTEM = (STOCKFULL_MANAGEMENT || STOCKEMPTY_MANAGEMENT)
                     /{decrementStockA/receive,
incrementStockB/send}.
```

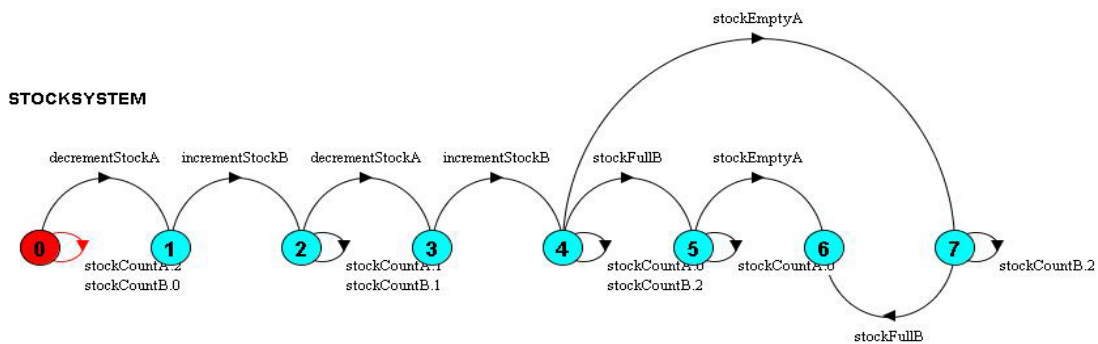

Figure 5. The LTS specifications for stock management

Liveness properties alone are not necessarily sufficient to describe the system limits and there is a need of certain invariants i.e. safety constructs to be maintained while executing.

**Safety - NOLOSS property**

Safety property NOLOSS of Carrier agent infers that there is no loss of stock during the carrier load, unload, and movements between the storehouses. To represent the LTS here with all its states, we have taken a mini-route with only three road partitions. The carrier is loaded and then the carrier is full, there is no loss of stock during the carrier agent's trajectory between storehouse A and B. Safety property specifies every trace that satisfies the property for a particular action alphabet. If the system produces traces that are not accepted by the property automata then a violation is detected during reachability analysis.

```
1    const N=2       // Number of carrier agents
2    const Min=0     // First(Load) road partition
3    const Max=3     // Last(Unload) road partition
4    property NOLOSS_Stock = (empty.loaded -> ONTHEWAY[1]),
5    ONTHEWAY[part:Min..Max] = (
6        when(part>Min && part<Max) full.moveto[part] -> ONTHEWAY[part+1]
7      | when(part==Max)            full.unloaded -> NOLOSS_Stock).
8    ||NOLOSS = (c[1..N]:NOLOSS_Stock).
```

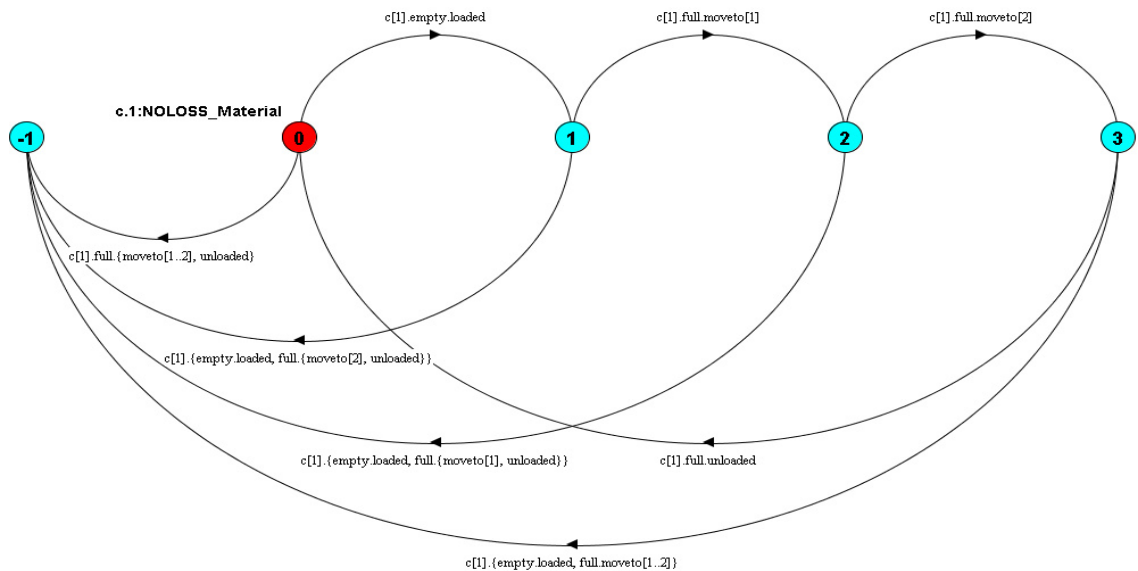

Figure 6. FSP specifications along with the generated LTS for NOLOSS

As shown by the LTS in figure-6 there is an error(-1) state to check the system safety, each and every action that is not compatible with the specifications results into the error state. As a result we have a system that is exhaustively verified.

## 5. Moving from requirement specification to requirement verification (Transformations from Gaia specifications to LTS specifications)

For the formal verification of safety and liveness properties of requirement specifications, the Gaia specifications have been transformed to LTS specifications. This whole process is done by hand. The Gaia role model liveness and safety properties along with the organizational rules are specified in the form of finite automates for verification.

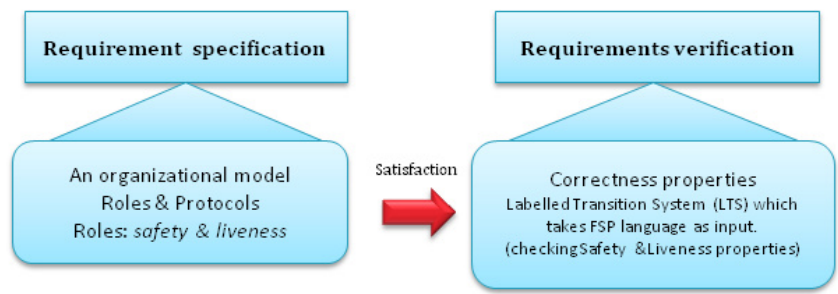

Figure-7: Moving from requirement specification to requirement verification

www.iiste.org

Table-3: Transformation (i.e. satisfaction relation) from Gaia based requirements to LTS based verification

| |
|---|
| Gaia: specifications have been founded on organizational abstractions, with different layers of abstractions represented by their corresponding models.<br>LTS: language is based on process algebra. In LTS agents, environment, organizational rules, agents role model, agent's protocols all are represented by a labelled transition system with processes, states, and actions. There are processes working in parallel with synchronization between them by action sharing. |
| **Environment**<br>Gaia: the environment is specified in terms of environmental model. It is an abstract, computational representation of the environment in which agents are situated. The environment is treated in terms of abstract computational resources e.g. variables or tuples, made available to the agents for sensing (reading their values), for effecting (changing their values) or for consuming (extracting them from the environment). So the environmental model is a list of resources; each associated with a symbolic name and characterized by the type of actions that the agents can perform on it.<br>For example:<br>***reads*** signNumber  *// readable resource of the environment*<br>***reads*** parking_flag  *// another readable resource*<br>***changes*** position  *// a variable that can also be changed by the agent*<br>A multi-agent system is always situated in some environment and it is considered the primary abstraction during the analysis and design phases. It is important to identify what should be characterized as part of environment, to identify the environmental resources that agents can effectively sense and effect, to identify how an agent should perceive its environment. The environment of a multi-agent system should not be implicitly assumed and its characteristics must be identified, modelled, and possibly shaped to meet application-specific purposes.<br>LTS: environment is also specified in terms of processes, states, and actions. The environment has to be translated into LTS. The physical environment becomes processes working concurrently with each process having one or more actions. |
| **Agents**<br>Gaia: agents are specified in terms of models. Role, Interaction, Agent, Services, and Acquaintance models. These models represent agents at various layers of abstractions.<br>LTS: agents along with their roles are expressed in the form of finite state processes. An agent is a single process or a concurrent execution of one or more processes with each process having one or more actions.<br>For example: Consider a stock manager agent which manages the count of stocks in a storehouse. It counts the stock and then informs the loader agent or un-loader agent about the stock count. It performs three actions countStock, informLoader, and informUnloader<br>       **Stock_manager = (   countStock -> Stock_manager**<br>                        **\| informLoader -> Stock_manager**<br>                        **\| informUnloader -> Stock_manager**<br>       **).**<br>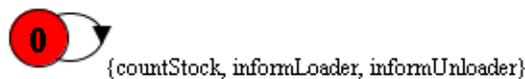<br>Consider a loader agent which loads the carrier with material. Below we have represented it as a sequence of actions<br>**LOADER = (waitForEmptyCarrier -> waitForStoreManagerToDeliver ->  loadCarrier -> LOADER).**<br>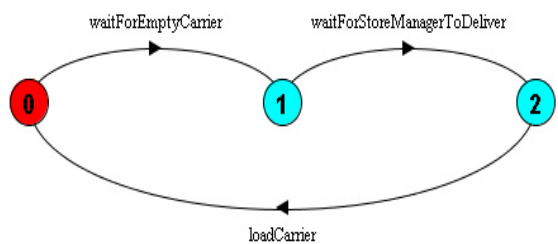 |

**Organizational rules**
Gaia: a well-defined syntax of organizational rules.
LTS: organizational rules can also be specified in the form of LTS with the same FSP language.

**Role model**
Gaia: we have been using different syntaxes to represent liveness and safety properties.
Liveness property is represented by a regular expression.
We have represented the safety property by the first-order predicate logic.
For example: Consider the safety property of an empty carrier agent moving on a road which has been broken down into discrete partitions. Each road partition having a distinct sign number. This property states that if the carrier agent is empty than it can move forward (to the next road part) and can also move back (to the previous road part)

**(is_Empty(c) ∧ can_movetoNext(sn)) ∨ (is_Empty(c) ∧ can_movetoPrevious(sn))**
where c is for carrier and sn for the sign number

LTS: The safety and liveness both are represented as a labelled transition system with processes, states, and actions. To represent a safety property the keyword property is used.
For example: Consider the same safety property of an empty carrier agent moving on a road which has 3 parts. We have taken 3 parts to represent each and every state. This property states that if the carrier agent is empty than it can move forward (to the next road part) and can also move back (to the previous road part)

```
const Min=1 /// First road partition
const Max=3 /// last road partition
property Empty_Carrier = (empty.start -> ROAD[1]),
ROAD[part:Min..Max] = (
      when(part>=Min && part<Max)    empty.movetoNext[part+1] -> ROAD[part+1]
    | when(part>Min && part<=Max)    empty.movetoPrevious[part-1] -> ROAD[part-1]
    | when(part==Max)                empty.stop -> Empty_Carrier).
```

Here below is the LTS with all its states

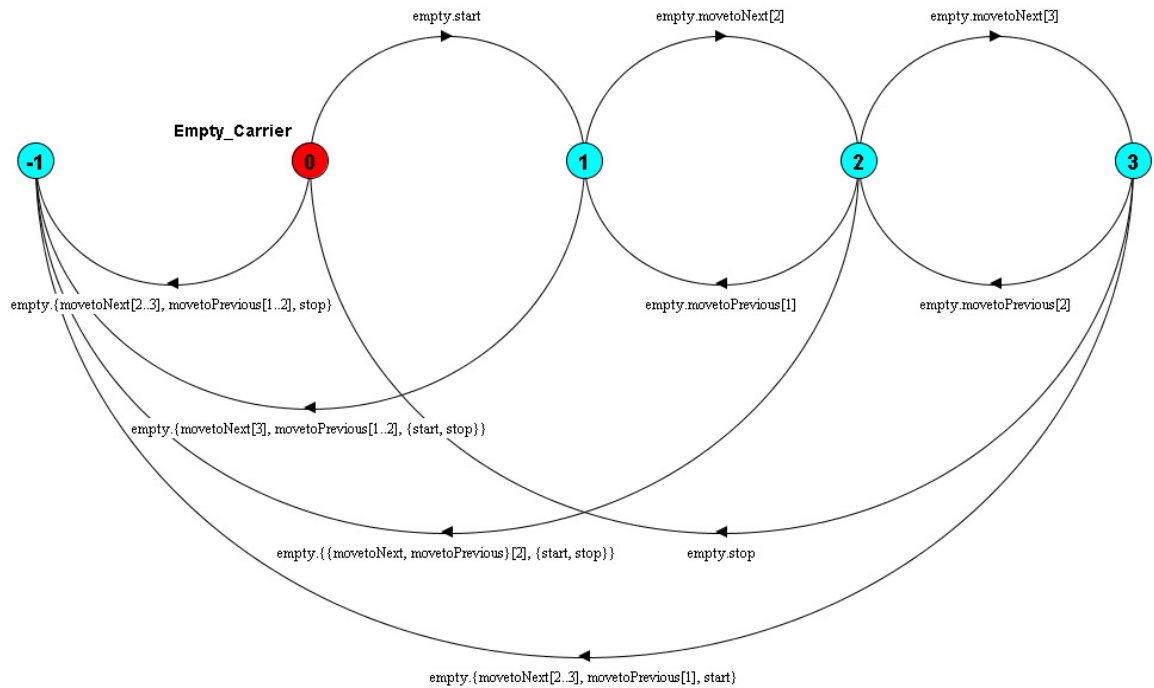

**Safety and Liveness properties**
Gaia: The safety and liveness properties are represented in terms of atomic components protocols and activities.
An *activity* is a unit of action that the agent may perform, which does not involve interaction with any other agent. *Protocols* are activities that do require interaction with other agents.
For example:
*Property NOLOSS*:
It infers that there is no loss of material during the carrier load, unload, and movements between the storehouses.

| |
|---|
| **carrier_isfull ∧ (material_loss = 0) ∧ (carrier_position = loadsign)** <br> ∨ <br> **carrier_isfull ∧ (material_loss = 0) ∧ (carrier_position = unloadsign)** <br> ∨ <br> **carrier_isfull ∧ (material_loss = 0) ∧ (carrier_position = sign)** <br> where loadsign is loader position, unloadsign is unloader position, sign is any road partition of the road. <br> LTS: the safety and liveness properties are expressed by processes and actions. There is no concept of protocols and activities. Process is the sequential execution of one or more actions. There may be multiple processes executing in parallel with synchronization between processes by action sharing (i.e. with the same name actions). <br> For example: Property NOLOSS expressed in FSP along with LTS in section 4.5. |
| Gaia: liveness expression dot(.) represent the sequence <br> x.y   indicates x followed by y <br> where x and y is a protocol or an activity <br> LTS: `x -> y`  action x followed by action y |
| Gaia: x \| y <br> choice between x and y where x and y are sequences of combination of protocols and activities <br><br> LTS: `X | Y` <br> choice between two processes X and Y. Each process is a combination of one or more actions. |
| Gaia:  x*   x occurs 0 or more times <br> LTS: for loop or repetition there is recursion. <br> `X_star = (stop -> STOP | x -> X_star).` |
| Gaia:   $x^+$   x occurs 1 or more times <br> LTS:  `X_plus = (x -> STOP | x -> X_plus).` |
| Gaia:   $x^w$   x occurs infinitely often <br> LTS:  `X_omega = (x -> X_omega).` |
| Gaia: [x]  x is optional <br> LTS: There is no specific operator for optional. |

Gaia: `x||y` x and y interleaved
There is no concept of shared actions. There are protocols that define the interaction between two agents.
LTS:
`||` construct is used for parallelism (concurrent execution)
`||` is used between processes; each process has one or more actions. So if each process has a single action then the two actions of two processes should be parallel.
`||` is also used as a composition i.e. composition of two or more processes

**Example**
A process X having single action    x_action
`X = (x_action -> X).`

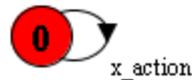

A process Y having a single action    y_action
`Y = (y_action -> Y).`

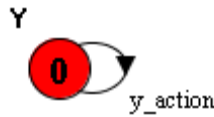

A process Z which is a parallel composition of the two processes X and Y. Therefore each action of each process, x_action in the case of process X and y_action in the case of process Y are executing in parallel i.e. concurrently
`||Z = (X || Y).`

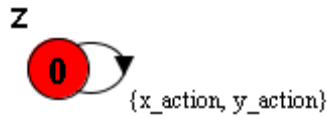

**Protocols and Interactions**
Gaia: The protocols are activities that require interaction with other agents. Therefore protocols play the role of connection between agent roles.
LTS: Interactions are modelled using shared actions. In a composition of processes, the actions that are common between processes are called shared. A shared action is executed at the same time by all processes that participate in the shared action.
`MAKER = (make -> ready -> MAKER).`

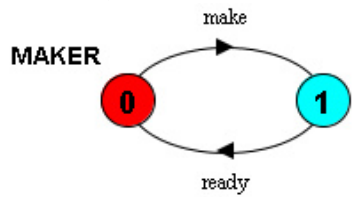

`USER = (ready -> use -> USER).`

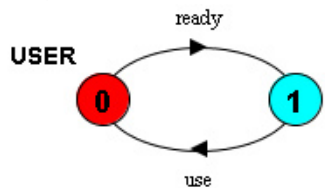

`||MAKER_USER = (MAKER || USER).`

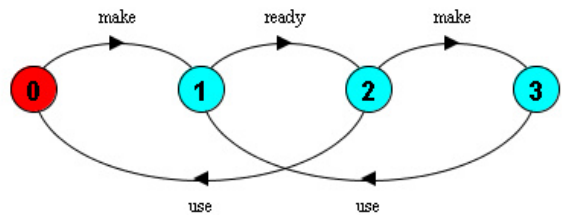

Here ready is an action that is the shared action between the processes MAKER and USER. Ready synchronizes the two processes.

**Protocols**
Gaia: There are protocols that define the interaction between different roles.
For example: Consider the Unload role.
Unload role is the role of the loader/unloader unloading the loaded carriers. waitforUnloading is a protocol between Move_full and Unload role.

*Unload role liveness property:*
**Unload** = *WaitForLoadedCarrier*.unloadCarrier
*WaitForLoadedCarrier* = (waitForCarrierPresence)+.WaitforUnloading

**Protocol** WaitforUnloading**:**

| WaitforUnloading | | sign_number |
|---|---|---|
| Move_full | Unload | |
| Wait for the Un-loader | | position |

LTS: it has shared actions. These shared actions define the interaction between parties.

**Services**
Gaia: There is a services model with the service name, input, output, pre-condition, and post-condition of each service.
LTS: A service is represented as a combination of processes executing concurrently with each process having one or more actions.

## 6. Lessons learned and Conclusion

Gaia's role model defines the behavior and responsibilities that are liveness and safety properties. Transformation from Gaia to FSP plays a key role for the formal verification of our system. In this paper we have not detailed the transformation process and its semantics but we would be presenting it in our future work.

Gaia method has a concrete syntax to express properties, is suitable to model behaviors and is applicable to a wide range of multi-agent systems but it does not provide constructs for the formal verification. Therefore we have to translate the Gaia concepts into FSP specifications. With these formal techniques and methods we studied the features of formal verification and property checking for multi-agent systems. There is a need for the development of a clear method, based on formal verification and organizational abstractions, for the analysis and design of multi-agent systems specifications. For the simulation of our case study example we are using service-based architecture Microsoft Robotics Developer Studio [MRDC, 2007].

The objective is to devise multi-agent systems based on formal methods that assure correctness. Multi-agent systems have concurrency, synchronization and deadlock issues to be handled and it's suitable to use formal development methods with organizational structure, appropriate set of agent abstractions and formal verification methods for checking the correctness of the system.

## References

[Wooldridge and Jennings, 1995] Wooldridge, M. and Jennings, N. R.: *Intelligent agents: Theory and practice*. Knowledge Engineering Review, 10(2):115-152, 1995.

[Jennings et al., 1998] Jennings, N. R., Sycara, K. and Wooldridge, M.: *A Roadmap of Agent Research and Development*. Int. Journal of Autonomous Agents and Multi-Agent Systems, 1 (1). pp. 7-38, 1998.

[Zambonelli et al., 2003] Zambonelli, F., Jennings, N. R., and Wooldridge, M.: *Developing Multiagent Systems: The Gaia Methodology*. ACM Transactions on Software Engineering and Methodology, 12(3):317-370, 2003.

[Wooldridge et al., 2000] Wooldridge, M., Jennings, N. R. and Kinny, D.: *The Gaia Methodology for Agent-Oriented Analysis and Design*. Autonomous Agents and Multi-Agent Systems, 3, 285-312, 2000

[Magee and Kramer, 2006] Magee, J. and Kramer, J.: *Concurrency: State Models and Java Programs*. John Wiley and Sons, 2nd edition, 2006.

[Giunchiglia et al., 2002] Giunchiglia, F., Mylopoulos. J., Perini. A.: *The tropos software development methodology: processes, models and diagrams*, Proceedings of the first international joint conference on Autonomous agents and multiagent systems: part 1, July 15-19, 2002, Bologna, Italy

[MRDC, 2007] Microsoft Robotics Developer Center: *http://msdn.microsoft.com/en-us/robotics/default.aspx*, 2007

[Abrial, 1996] Abrial, J. R.: *The B-Book*. Cambridge University Press, Cambridge, U.K. 1996.

[Clarke et al., 2000] Clarke, E., Grumberg, O. and Peled, D.: *Model Checking*. MIT Press, 2000.

[Berard et al., 2001] Berard, B., Bidoit, M., Finkel, A., Laroussinie, F., Petit, A., Petrucci, L., Schnoebelen, P. and McKenzie, P.: *Systems and Software Verification: Model-Checking Techniques and Tools*. Springer-Verlag, 2001.

[Wooldridge, 2000] Wooldridge, M.: *Reasoning about Rational Agents. Intelligent Robot and Autonomous Agents Series*. MIT Press, Cambridge, MA. 2000.

[Giannakopoulou et al., 1999] Giannakopoulou, D., Magee, J. and Kramer. J.,: *Fairness and priority in progress property analysis*. Technical report, Department of Computing, Imperial College of Science, Technology and Medicine, 180 Queens Gate, London SW7 2BZ, UK, 1999.

**Dr. Nadeem Akhtar** is working as an Assistant Professor at the Department of Computer Science & IT, The Islamia University of Bahawalpur. He has a PhD from the research Laboratory VALORIA of Computer Science, University of South Brittany (UBS), France. His research areas are formal specification, formal architecture, and service-oriented architecture for robotics.